\documentclass[journal]{IEEEtran}

\IEEEoverridecommandlockouts
\usepackage{cite}
\usepackage{amsmath,amssymb,amsfonts}
\usepackage{algorithm}
\usepackage{algpseudocode}
\usepackage{graphicx}
\usepackage{textcomp}
\usepackage{xcolor}
\usepackage{xspace}
\usepackage{tabularx}
\usepackage{enumitem}
\usepackage{listings}

\begin{document}

\graphicspath{{figs/}}

\newcommand{\jinting}[1]    {\textbf{\textcolor{red}{[jinting] #1}}}
\newcommand{\jintingd}[1]    {\textbf{\textcolor{green}{[jinting] #1}}}
\newcommand{\tengfei}[1]    {\textbf{\textcolor{purple}{[tengfei] #1}}}
\newcommand{\monaas}        {Monaas\xspace}

\title{\monaas: Mobile Node as a Service for TSCH-based Industrial IoT Networks}

\author{
    {Jinting Liu, Jingwei Li, Tengfei Chang\textsuperscript{*}}
    \thanks{The authors are with Information Hub of The Hong Kong University of Science and Technology (Guangzhou), Guangdong, China (e-mail: jliu738@connect.hkust-gz.edu.cn; jli186@connect.hkust-gz.edu.cn; tengfeichang@hkust-gz.edu.cn).}
    \thanks{*Corresponding Author}
}

\maketitle

\begin{abstract}

The Time-Slotted Channel Hopping (TSCH) mode of IEEE802.15.4 standard provides ultra high end-to-end reliability and low-power consumption for application in field of Industrial Internet of Things (IIoT).
With the evolving of Industrial 4.0, 
    dynamic and bursty tasks with varied Quality of Service (QoS); 
    effective management and utilization of growing number of mobile equipments 
become two major challenges for network solutions.
The existing TSCH-based networks lack of a system framework design to handle these challenges.
In this paper, we propose a novel, service-oriented, and hierarchical IoT network architecture named Mobile Node as a Service (\monaas).
\monaas aims to systematically manage and schedule mobile nodes as on-demand, elastic resources through a new architectural design and protocol mechanisms. 
Its core features include 
    a hierarchical architecture to balance global coordination with local autonomy, 
    task-driven scheduling for proactive resource allocation, and 
    an on-demand mobile resource integration mechanism. 
The feasibility and potential of the \monaas link layer mechanisms are validated through implementation and performance evaluation on an nRF52840 hardware testbed, demonstrating its potential advantages in specific scenarios.
On a physical nRF52840 testbed, \monaas consistently achieved a Task Completion Rate (TCR) above 98\% for high-priority tasks under bursty traffic and link degradation, whereas all representative baselines (Static TSCH, 6TiSCH Minimal, OST, FTS-SDN) remained below 40\%. Moreover, its on-demand mobile resource integration activated services in 1.2 s, at least 65\% faster than SDN (3.5 s) and OST/6TiSCH ($>$ 5.8 s).

\end{abstract}


\section{Introduction}
\label{sec:intro}

 adopted in industrial Internet of Things (IIoT) applications due to its ability to provide high communication reliability 
    while maintaining ultra-low power consumption~\cite{dujovne146tisch}.  
However, the evolution toward Industry~4.0 introduces fundamentally different system requirements, 
    including highly event-driven traffic patterns, 
    stronger spatial and temporal correlations, and 
    increasing demands for low-latency on-demand communications~\cite{lu2017industry}.  
In parallel, modern industrial environments are no longer composed solely of static low-power sensor nodes, but 
    increasingly integrate mobile agents such as automated guided vehicles (AGVs) and aerial drones equipped with advanced sensors, actuators, and substantial onboard computing capabilities\cite{park20multi}.  
These mobile and resource-rich devices represent a significant pool of untapped, 
    on-demand system resources that are largely overlooked by existing TSCH protocols and scheduling research.

In contemporary industrial management of warehouses and factories, 
    such as asset tracking and environmental conditions, 
    resource-constrained and low-cost sensing devices are widely deployed to continuously monitor basic operational states.
Meanwhile, mobile platforms including automated guided vehicles (AGVs) and aerial drones are increasingly employed to perform routine inspection, inventory verification, and maintenance-related tasks.
Existing TSCH research has predominantly focused on preserving network connectivity and schedule stability in the presence of node mobility~\cite{duquennoy2017contiki,ngo2019user},  
    typically modeling mobile entities as sources of performance perturbation whose impact must be minimized or compensated.  
However, such approaches largely lack a semantic understanding that bridges high-level task requirements with the heterogeneous sensing, actuation, communication, and computation capabilities of available mobile resources.  
As a result, current TSCH-based systems miss significant opportunities to enable elastic, task-driven resource allocation that can fully exploit the on-demand potential of mobile industrial agents.

To address this gap, we propose \textsc{MonAAS}, a Mobile Node as a Service framework for TSCH-based industrial networks.  
\textsc{MonAAS} introduces a service-oriented scheduling abstraction that elevates mobile nodes from passive network participants to active service providers.  
By explicitly exposing node capabilities to the scheduling layer, 
    the framework enables scheduling decisions that are aware of sensing, actuation, communication, and computation resources available in the network. 
In contrast to conventional TSCH designs, \textsc{MonAAS} incorporates a recruitment mechanism that allows the network to dynamically request mobile devices to provide specific services.  
This mechanism enables the elastic involvement of mobile agents, 
    such as AGVs and drones, in response to evolving system demands.  
As a result, mobile resources can be opportunistically activated and coordinated rather than being statically embedded into the schedule.
Furthermore, \textsc{MonAAS} supports on-demand scheduling driven by the time and location of detected events.
Scheduling decisions are therefore aligned with when and where services are required.  
This event-aware design enables timely and resource-efficient responses while preserving the reliability guarantees of TSCH-based communication.

This work focuses on the MAC layer implementation of the \monaas framework, providing the foundational scheduling and resource orchestration mechanisms that bridge high-level task semantics with low-level TSCH communication primitives.
The main contributions of this paper are summarized as follows:

\begin{itemize} [leftmargin=*]

    \item We propose a \textbf{service-oriented scheduling framework} that models mobile nodes as elastic service resources and enables task-driven capability matching with execution-zone-aware scheduling.
    
    \item We design an \textbf{on-demand mobile resource integration mechanism} that proactively discovers, recruits, and temporarily integrates mobile nodes based on real-time task demands rather than reactive connectivity handling.
    
    \item We develop a \textbf{hierarchical network architecture for \monaas} that balances global coordination with regional autonomy, allowing local Leaders to make timely decisions while preserving system-wide scalability.
    
    \item We implement and evaluate the \textbf{core \monaas logic at the MAC layer} on a physical nRF52840-DK testbed, demonstrating significant improvements in task completion rate (over 98\% versus below 40\%) and mobile resource integration latency (1.2\,s versus over 3.5\,s).
\end{itemize}

The remainder of this paper is organized as follows.  
Section~\ref{sec:related} reviews existing mobility management approaches in TSCH networks and highlights the fundamental paradigm gap addressed by \monaas.  
Section~\ref{sec:monnas} presents the \monaas system architecture, including its hierarchical design, core mechanisms, and task-driven resource orchestration strategy.  
Section~\ref{sec:setup} describes the experimental methodology, hardware testbed based on nRF52840-DK devices, benchmark configurations, and evaluation scenarios.  
Section~\ref{sec:analysis} analyzes the experimental results, demonstrating the effectiveness of \monaas in dynamic environments and mobile resource integration.  
Section~\ref{sec:limitation} discusses current limitations and outlines future.

\section{Related Work}
\label{sec:related}

This section reviews related work on TSCH scheduling with a particular emphasis on mobility support in industrial networks.  
We first provide a concise overview of TSCH scheduling fundamentals, including the slotframe structure and matrix-based scheduling model.  
We then review existing TSCH scheduling mechanisms proposed for mobile scenarios, showing that 
    most prior approaches treat mobility primarily as a source of network disruption to be mitigated and 
    lack semantic interpretation of mobile devices capabilities.

\subsection{Background Technology: How TSCH Works}

\begin{figure*}[htbp]
    \centerline{\includegraphics[width=0.9\textwidth]{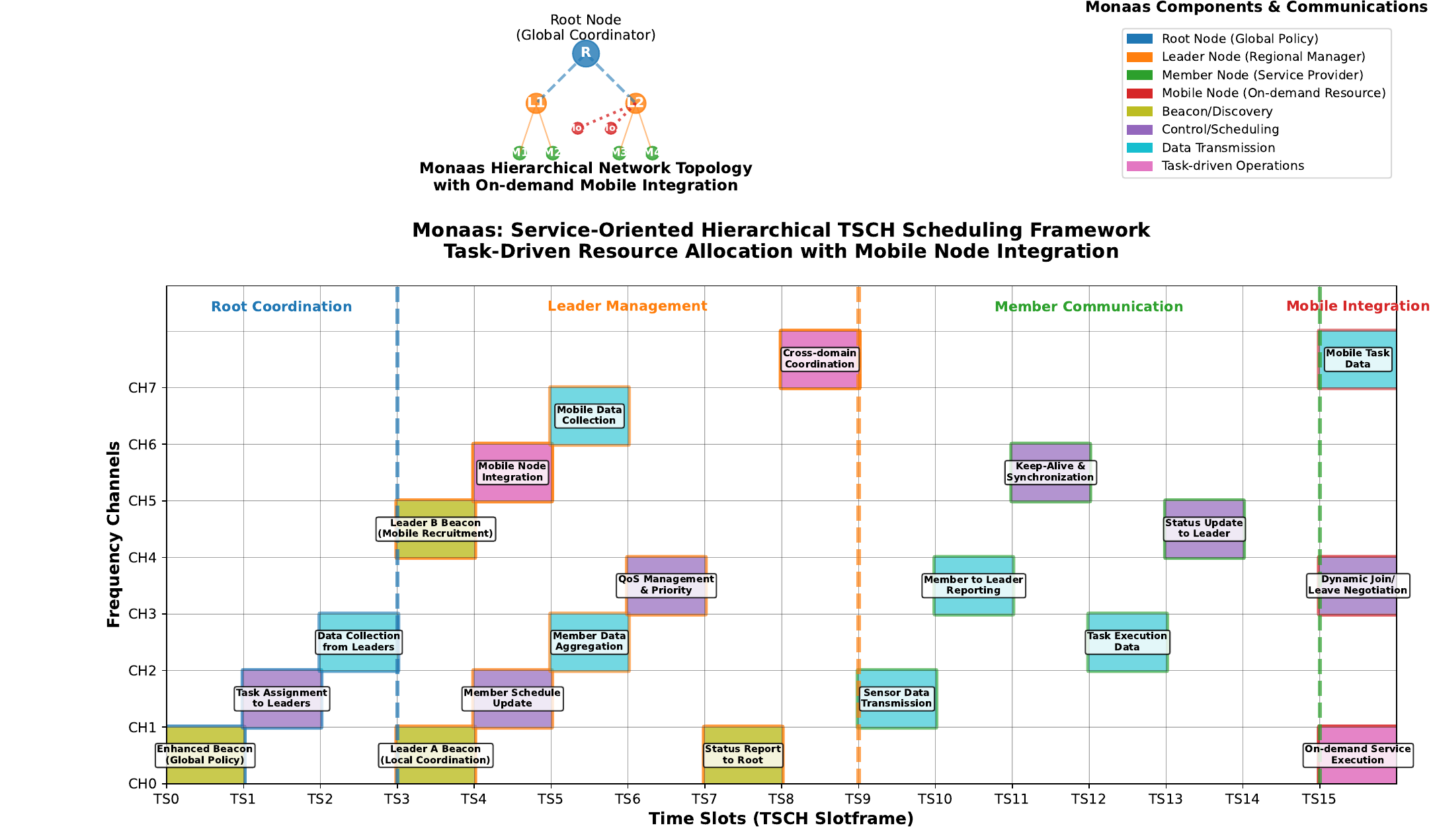}}
    \caption{
    An illustrative TSCH timing diagram showing communication tasks performed by different nodes (Root, Leader, Member, Mobile) across different time slots and channels.
    }
    \label{fig:tsch}
\end{figure*}

IEEE 802.15.4 TSCH is a MAC layer protocol designed for high-reliability, low-power IIoT applications. Its fundamental principle is the virtualization of the time and frequency domains into a structured \textbf{schedule matrix}, which governs all network communications. This matrix is a grid where time is divided into \textbf{Time Slots}, and the radio spectrum is partitioned into multiple \textbf{Channels}. A specific (slot, channel) combination in this matrix is called a \textbf{cell}, which represents the fundamental unit of \textbf{network resource} for communication. As illustrated in Figure \ref{fig:tsch}, nodes are allocated specific cells to transmit or receive data, ensuring contention-free communication. For example, a Leader might use a dedicated broadcast cell to send commands, while a Member uses a unicast cell to report data to its Leader.

In this paper, we strictly distinguish between two types of resources: 1) \textbf{Network Resources}, which refer to the TSCH cells used for communication, and 2) \textbf{Service Resources}, which denote the physical or computational capabilities of a node (e.g., sensing, actuation, data processing). This distinction is critical to understanding why existing approaches fall short: they excel at allocating network resources but lack mechanisms to discover, match, and orchestrate service resources based on task requirements with spatial and temporal constraints.

In a standard 6TiSCH network, the operational flow typically begins with a new node listening for \textbf{Enhanced Beacons (EBs)} from existing nodes. These beacons contain network synchronization and schedule information. Upon receiving an EB, the new node synchronizes and can join the network. It then negotiates with its parent node (e.g., using the 6P protocol) to acquire dedicated cells for upstream data transfer. This process establishes a reliable, multi-hop path to the network root.
This scheduled approach provides two key advantages:
\begin{itemize}[leftmargin=*]
    \item \textbf{High Energy Efficiency:} Nodes activate their radios only during their assigned cells and remain in a low-power sleep state otherwise, drastically reducing energy consumption.
    \item \textbf{High Reliability:} By hopping across different channels in successive time slots, TSCH effectively mitigates persistent interference and multipath fading, ensuring robust communication links.
\end{itemize}
Having introduced the fundamental principles of TSCH, the remainder of this section will review related work from several key dimensions to highlight the context and innovations of the \monaas architecture.

\subsection{Critique of Mobility Handling in TSCH: A Connectivity-Centric View}

A review of existing TSCH literature reveals a dominant design paradigm in which mobility is primarily treated as a connectivity challenge to be mitigated rather than as a service opportunity to be exploited.  
This perspective directly reflects the original problem domain of TSCH protocols, which were designed to support reliable and energy-efficient communication in largely static industrial sensor networks.  

\subsubsection{Mobility as a Disruption to Connectivity}

Foundational TSCH scheduling protocols, including centralized approaches such as TASA~\cite{jin2016centralized} and distributed solutions such as Orchestra~\cite{duquennoy2015orchestra} and ALICE~\cite{kim2019alice}, were engineered for low-power and quasi-static TSCH-based wireless sensor networks.  
Their primary objective is to guarantee extreme reliability and energy efficiency under stable network conditions.
Within this design context, node mobility is inherently viewed as a disruptive event that threatens established communication paths and schedule consistency.  
As a result, the mobility-handling mechanisms developed in these systems are fundamentally defensive in nature.  

Centralized SDN-based approaches~\cite{bello2020sdn,pettorali2024mobility} exemplify this perspective by focusing on preserving end-to-end path integrity under mobility.  
A central controller tracks mobile nodes and recomputes schedules to ensure that their data can continue to reach the network root.  
The underlying objective is to preserve the connectivity of the moving node.  
Similarly, distributed schemes such as OST~\cite{jeong2020ost} react to mobility by detecting link failures and initiating neighbor discovery to locally repair the topology.  
Although differing in control structure, both approaches pursue the same goal of reactive link maintenance rather than proactive resource utilization.  
They address the question of how to keep a moving node connected, rather than what functionality that node can provide.  

Beyond these baseline designs, several works explicitly address mobility and quality-of-service in TSCH and 6TiSCH networks.  
The mobility-aware framework proposed by Al-Nidawi and Kemp~\cite{alnidawi2015mobility} improves handover robustness by accounting for link dynamics, but remains focused on connectivity maintenance during transitions.  
The MSU-TSCH family~\cite{jerbi2023msu,jerbi2024enhanced} updates local schedules in response to detected mobility events, enabling faster route and cell recovery through post-event adaptation.  
Tavallaie et al.~\cite{tavallaie2021design} propose a traffic-aware scheduler for mobile 6TiSCH networks, improving resource allocation under mobility but still driven by traffic metrics rather than task semantics.  
From a control-plane perspective, SDN-TSCH~\cite{veisi2022sdn} introduces traffic isolation to preserve per-flow guarantees, while multi-objective QoS optimization~\cite{vatankhah2024qos} balances delay, reliability, and energy efficiency.  
While these approaches advance mobility robustness and QoS assurance, they remain fundamentally connectivity- and flow-centric.  
Mobile nodes are still treated as moving endpoints whose connections must be preserved, rather than as elastic resources whose capabilities can be proactively orchestrated.  

\subsubsection{The Semantic Gap in Mobility Awareness}

This connectivity-centric design philosophy creates a fundamental semantic gap between MAC-layer events and their application-layer significance.  
Consider an industrial scenario in which an AGV equipped with gas sensors and high-definition cameras enters a zone experiencing a chemical leak.  
At the MAC layer, this event is merely observed as the detection of a new neighboring node through packet reception.  
At the application level, however, the system has gained access to critical sensing and inspection capabilities precisely where they are urgently required.  
Existing TSCH protocols lack mechanisms to bridge this gap, as they interpret mobility in terms of addresses and links rather than services and capabilities.  

Current mobility management schemes~\cite{jerbi2023msu, duquennoy2017contiki} operate entirely at the network level and provide no abstraction to associate node identities with their functional capabilities.  
As a result, the network becomes aware of new MAC addresses while remaining blind to the resources those addresses represent.  
Without such semantic translation, the system cannot perform task-driven decisions or proactively recruit suitable mobile agents.  
From the protocol perspective, an AGV is indistinguishable from any other node.  
This disconnect between low-level network events and high-level service requirements fundamentally limits the role of mobile nodes in TSCH systems.  
It prevents their transformation from unstable endpoints into on-demand service providers~\cite{ngo2019user}.  
\monaas is explicitly designed to bridge this semantic gap.   

\subsection{Summary}

Based on the literature review, existing TSCH scheduling approaches exhibit three fundamental limitations.  
First, they adopt a connectivity-centric worldview that treats mobile nodes primarily as connectivity burdens to be managed rather than as service providers to be exploited.  
Second, they lack task semantic awareness, resulting in no effective mapping from high-level task requirements, such as required capabilities, execution zones or time windows, and QoS constraints, to concrete scheduling decisions.  
Third, they provide no explicit service resource orchestration mechanisms to discover, match, and coordinate heterogeneous node capabilities in response to dynamic task demands.  

To systematically address these limitations, three key research questions must be answered.  
\begin{itemize}
    \item \textbf{Architectural Design:} How can a hybrid network architecture be designed to maintain global policy consistency while enabling regional autonomy for rapid response to local dynamics, particularly the joining and leaving of mobile nodes?
    
    \item \textbf{Task--Resource Matching:} How can network scheduling decisions move beyond traditional traffic-driven models to incorporate task semantics, including service type, priority, and QoS requirements, together with heterogeneous node capabilities?
    
    \item \textbf{Mobile Resource Integration:} How can protocol mechanisms be designed to proactively discover, recruit, integrate, and schedule mobile nodes based on real-time task demands, thereby enabling elastic, on-demand expansion of network service capacity?
\end{itemize}

The proposed \monaas framework addresses these questions through a hierarchical architecture, a semantics-aware task scheduling mechanism, and a mobile resource integrated protocol design.

\section{\monaas: A Mobile Resource-Oriented System Framework}
\label{sec:monnas}

\subsection{System Architecture Overview}

\begin{figure}[t]
    \centerline{\includegraphics[width=\columnwidth]{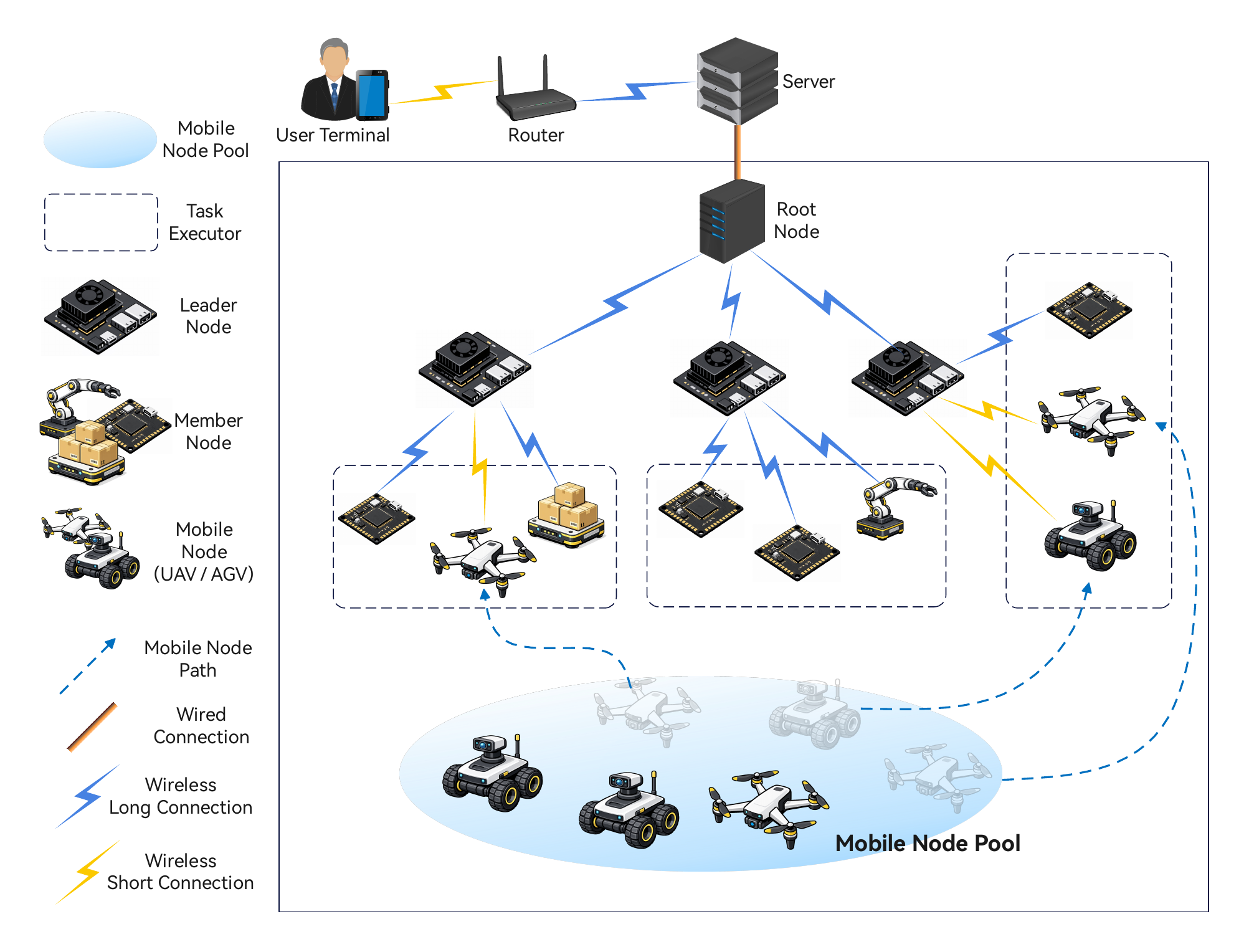}}
    \caption{\monaas hierarchical architecture showing the four key components: Root Node (global coordinator), Leader Nodes (regional managers), Member Nodes (static service providers), and Mobile Nodes (elastic resources).}
    \label{fig:arch}
\end{figure}

Fig.~\ref{fig:arch} illustrates the overall architecture of the \monaas framework and its interaction with heterogeneous network entities.  
The system adopts a hierarchical design composed of a Root node, multiple Leader nodes, static Member nodes, and mobile nodes with diverse capabilities.  

The Root node (R) resides at the top of the hierarchy and is responsible for maintaining global policies, network-wide task definitions, and long-term scheduling constraints.  
It communicates with Leader nodes through reliable backbone links and does not directly participate in fine-grained task execution or local mobility handling.  

Leader nodes (L) act as regional controllers that manage localized network segments.  
Each Leader maintains awareness of local Member nodes (M) and dynamically available mobile nodes (Mo), including their capabilities, status, and recent activity.  
Leaders are responsible for translating high-level task requirements received from the Root into region-specific scheduling decisions and recruitment actions.  

Member nodes represent static, resource-constrained TSCH devices that perform continuous sensing and reporting tasks.  
These nodes operate under deterministic TSCH schedules and form the stable backbone of the network.  

Mobile nodes (Mo) constitute an elastic pool of on-demand service resources.  
They may include AGVs, drones, or other mobile platforms equipped with advanced sensing, actuation, communication, and computation capabilities.  
Mobile nodes can dynamically join or leave Leader regions as they move, following the dotted mobility paths shown in the figure.  

When a task event occurs, the responsible Leader evaluates task semantics, such as required service type, execution location, and QoS constraints.  
Based on this information, the Leader selectively recruits suitable mobile nodes from the local or neighboring mobile pools and integrates them into the TSCH schedule (Sec.~\ref{sec:task}).  
This recruitment and integration process is realized through explicit protocol message exchanges, enabling temporary and task-driven participation of mobile resources(Sec.~\ref{sec:flow}).  

By separating global coordination at the Root from localized decision-making at Leaders, the architecture achieves both policy consistency and rapid responsiveness to mobility and dynamic task demands.  
This hierarchical and service-oriented design enables \monaas to transform mobile nodes from unstable endpoints into proactively orchestrated service providers.  

To formalize the architecture shown in Fig.~\ref{fig:arch}, we model a \monaas network as a set of nodes $N$, consisting of a single Root node $R$, a set of Leader nodes $L$, a set of static Member nodes $M$, and a pool of Mobile nodes $Mo$:

\begin{equation}
    N = \{R\} \cup L \cup M \cup Mo
\end{equation}

Each Leader node $l \in L$ manages a dynamic local domain $D_l(t)$, which includes its subordinate static Member nodes and the Mobile nodes currently associated with it:

\begin{equation}
\begin{aligned}
    D_l(t) ={}& \{ m \in M \mid \mathrm{parent}(m) = l \}  \cup \\
              & \{ mo \in Mo \mid \mathrm{associated}(mo, l, t) \}
\end{aligned}
\end{equation}

Here, $\mathrm{parent}(m)$ denotes the Leader node responsible for static node $m$, and $\mathrm{associated}(mo, l, t)$ indicates that mobile node $mo$ is associated with Leader $l$ at time $t$.

The overall architecture and the placement of Leader nodes are defined according to the physical deployment of the industrial environment.  
In practice, Leader roles and regional boundaries can be statically configured at deployment time or dynamically established using location-aware approach~\cite{pettorali24lasa}.
Static Member nodes are typically assigned to a specific Leader based on their deployment region.  
This association can be pre-configured or established through lightweight discovery and association mechanisms, such as enhanced beacon (EB) exchanges, similar to those used by mobile nodes.  
Once assigned, a static Member node generally remains associated with the same Leader for the lifetime of the deployment.

\subsection{Task-Driven Capability-Aware Scheduling}
\label{sec:task}

\monaas employs a task-driven, capability-aware scheduling policy that reserves network bandwidth based on both task requirements and node capabilities.  
In \monaas, a \textbf{Task} $T$ is formally defined as a tuple that encapsulates all information required for scheduling and execution:

\begin{equation}
\label{eq:task}
    T = \{ id, P, Q, C_{req}, Z_{target}, W_{target} \}
\end{equation}

Here, $id$ uniquely identifies the task.  
$P$ denotes the task priority, for example $P \in \{1, 2, 3, 4\}$, representing \text{Low}, \text{Medium}, \text{High}, \text{Critical}.  
$Q$ represents the Quality-of-Service (QoS) requirements, including constraints such as maximum tolerable latency $lat_{max}$ and minimum packet delivery ratio $pdr_{min}$.  
$C_{req}$ specifies the set of capabilities required to execute the task, such as $\{\text{gas\_sensor}, \text{high\_speed\_report}\}$.  
$Z_{target}$ defines the target execution zone, constraining where the task must be performed.  
$W_{target}$ defines the execution time window, constraining when the task should be executed and when allocated resources may be released.  

In addition to task definitions, \monaas characterizes each network node $n \in N$ using a set of runtime attributes.  
$C_n$ denotes the set of capabilities provided by node $n$.  
$S_n$ captures the real-time operational state of the node, such as remaining battery energy $b_n$.  
$load_n$ represents the current traffic load of node $n$, measured as the number of packets queued for transmission.

\begin{figure}[t]
    \centerline{\includegraphics[width=\columnwidth]{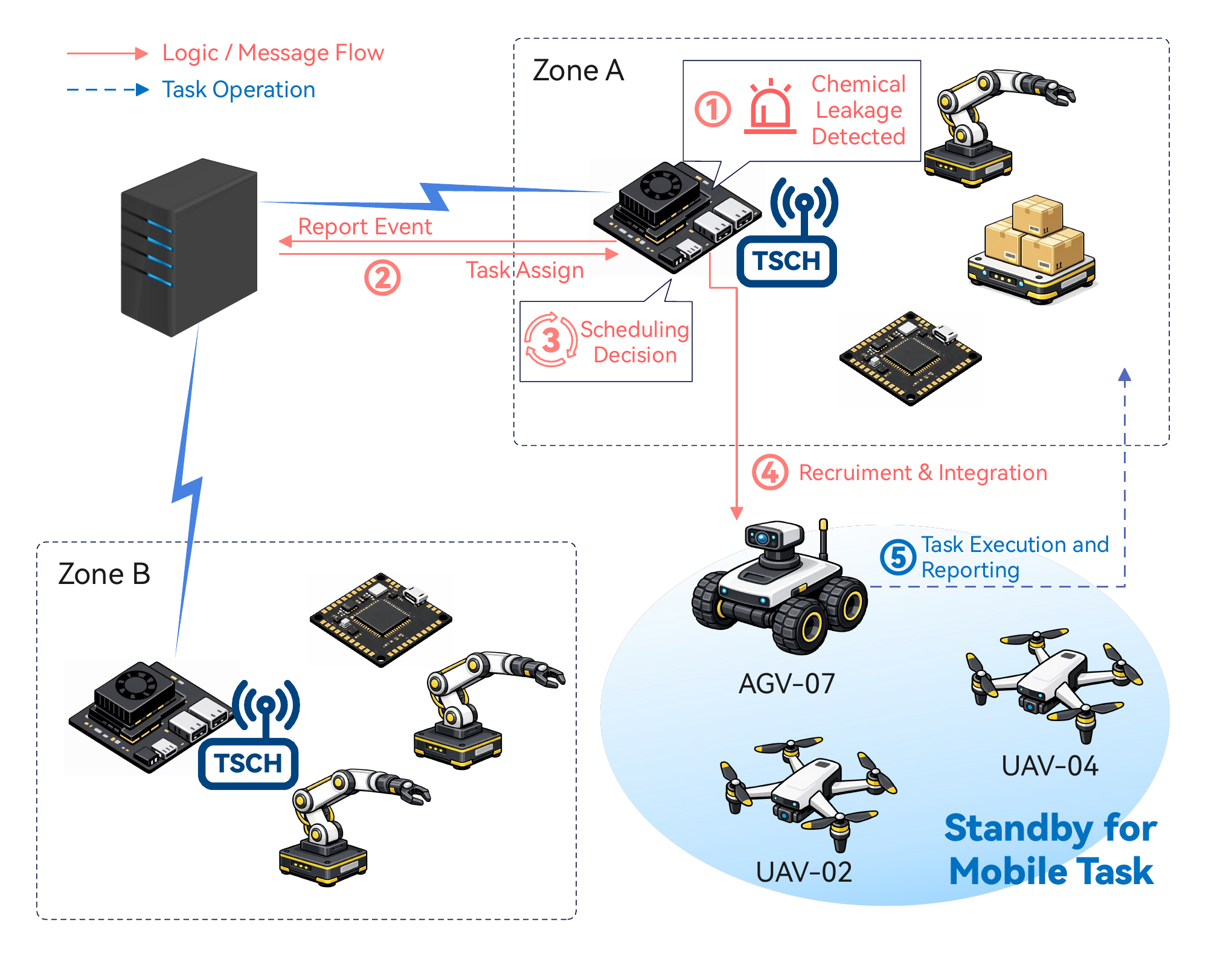}}
    \caption{Illustrative deployment of \monaas in a factory setting. The Root node serves as the central factory controller. Leader nodes manage specific zones (Zone A, B), overseeing static Member nodes (e.g., environmental sensors) and interacting with Mobile nodes (e.g., AGVs) that move between zones.}
    \label{fig:deployment_scenario}
\end{figure}

The translation from task requirements to a concrete number of TSCH timeslots is handled by a configurable policy function, \lstinline{EstimateSlots(T)}.  
This function converts high-level task semantics into MAC-layer resource requirements.  

For a task $T$, the required number of slots per slotframe is estimated as:

\begin{equation}
    \mathit{req\_slots}
    = \mathit{pkt\_sf} \cdot \mathit{retx} \cdot f_P
\end{equation}
where each term is derived from task attributes as follows.  

$\mathit{pkt\_sf}$ denotes the number of packets generated within one slotframe period and is defined as

\begin{equation}
    \mathit{pkt\_sf} = r_{T_{id}} \cdot T_{sf}
\end{equation}

where $r_{T_{id}}$ is the task-specific packet generation rate and $T_{sf}$ is the slotframe duration.  

$\mathit{retx}$ captures the retransmission factor required to satisfy the task’s reliability constraint and is computed as

\begin{equation}
    \mathit{retx} = \frac{Q.\mathit{pdr}_{\min}}{L_{\mathit{est}}}
\end{equation}

where $L_{\mathit{est}}$ denotes the estimated link delivery probability.
It usually can be monitored through Keep-Alive packets in TSCH networks by recording the number of transmission and number of being ACK'ed. 
When strict reliability is not required, a lower $Q.\mathit{pdr}_{\min}$ allows fewer retransmissions, reducing latency and resource consumption.  

$f_P$ is a priority-dependent scaling factor derived from the task priority $P$.  
Higher-priority tasks are assigned larger $f_P$ values to provide scheduling slack and reduce latency under contention.  

The final $\mathit{req\_slots}$ is then transformed into following equation 

\begin{equation}
\label{eq:slots}
    \mathit{req\_slots}
    = \left\lceil r_{T_{id}} \cdot T_{sf} \cdot \frac{Q.\mathit{pdr}_{\min}}{L_{\mathit{est}}} \cdot P \right\rceil
\end{equation}

The resulting $\mathit{req\_slots}$ is rounded to an integer and used by the Leader to allocate TSCH cells.
Each Leader’s timeslot resources are pre-assigned by the Root node and can be dynamically adjusted in response to runtime task demands.  
This approach avoids the complexity and signaling overhead associated with fully distributed timeslot negotiation among nodes, while still enabling elastic resource allocation.

At the global level, the Root node maintains a timeslot resource pool $R_{\mathit{global}}$ and pre-allocates baseline resources to each Leader according to historical traffic patterns and anticipated task demand.  
Each Leader node $l$ controls a local resource pool $R_l \subseteq R_{\mathit{global}}$ and independently allocates timeslots to tasks within its managed domain based on task requirements and local network conditions.  
When the local resource pool is insufficient to satisfy newly arriving tasks, the Leader directly requests additional timeslot resources from the Root, without engaging in negotiation with other Leaders.
Formally, the resource state of a Leader $l$ at time $t$ is defined as

\begin{equation}
    R_l(t) = R_{\mathit{base}}(l) \cup R_{\mathit{alloc}}(l,t)
\end{equation}

where $R_{\mathit{base}}(l)$ denotes the baseline resources pre-allocated by the Root, and $R_{\mathit{alloc}}(l,t)$ represents additional timeslots dynamically granted at time $t$.  

\begin{algorithm}[t]
\caption{Leader Node Task Processing and Resource Allocation}
\label{alg:leader_decision}
\begin{algorithmic}[1]
\State \textbf{Input:} Task $T$, Leader $l$, domain $D_l(t)$, resources $R_l(t)$
\State \textbf{Output:} Resource allocation decision and execution plan

\Statex \textit{--- Stage 1: Communication Resource Check ---}
\State $\mathit{req\_slots} \gets \textsc{EstimateSlots}(T)$
\If{$|R_l(t)| < \mathit{req\_slots}$}
    \State $\mathit{add\_needed} \gets \mathit{req\_slots} - |R_l(t)|$
    \State $\textsc{SendToRoot}(\textsc{REQ\_Slots}(l, \mathit{add\_needed}))$
    \State $\textsc{WaitForResponse}(\mathit{timeout})$
    \If{slots were granted}
        \State $R_l(t) \gets R_l(t) \cup \mathit{granted\_slots}$
    \Else
        \State \textbf{return} FAILURE \Comment{Request denied}
    \EndIf
\EndIf

\Statex \textit{--- Stage 2: Service Capability Check and Recruitment ---}
\State $\mathit{cap\_nodes} \gets \textsc{FindCapableNodes}(D_l(t), T.C_{req})$
\If{$|\mathit{cap\_nodes}| < \textsc{MinNodes}(T)$}
    \State $\mathit{need\_count} \gets \textsc{MinNodes}(T) - |\mathit{cap\_nodes}|$
    \State $\mathit{miss\_caps} \gets T.C_{req} \setminus (\bigcup \{C_n \mid n \in \mathit{cap\_nodes}\})$
    \State $\textsc{BroadcastBeacon}(T, \mathit{miss\_caps})$
    \State $\mathit{candidates} \gets \textsc{CollectRequests}(\mathit{timeout})$
    \State $\mathit{selected} \gets \textsc{SelectBest}(\mathit{candidates})$
    \ForAll{$mo \in \mathit{selected}$}
        \State $\textsc{GrantAccess}(mo, l)$
        \State $D_l(t+1) \gets D_l(t) \cup \{mo\}$
        \State $\mathit{cap\_nodes} \gets \mathit{cap\_nodes} \cup \{mo\}$
    \EndFor
\EndIf

\Statex \textit{--- Stage 3: Final Task Allocation ---}
\If{$|\mathit{cap\_nodes}| \geq \textsc{MinNodes}(T)$}
    \State $\mathit{sel\_nodes} \gets \textsc{SelectOptimal}(\mathit{cap\_nodes}, T)$
    \State $\mathit{slots} \gets \textsc{AllocateSlots}(\mathit{sel\_nodes}, T, R_l(t))$
    \State $\textsc{SendAssignment}(\mathit{sel\_nodes}, T, \mathit{slots})$
    \State \textbf{return} SUCCESS
\Else
    \State \textbf{return} FAILURE \Comment{Insufficient resources}
\EndIf
\end{algorithmic}
\end{algorithm}

Alg.~\ref{alg:leader_decision} describes the procedure executed by a Leader node to process a task request and perform task-driven resource allocation under the \monaas framework.  
The algorithm is triggered whenever a new task $T$ arrives at a Leader $l$ and operates on the Leader’s current domain $D_l(t)$ and local timeslot resource pool $R_l(t)$.  

In Stage~1, the Leader first translates the task requirements into concrete MAC-layer demand by computing the required number of timeslots $\mathit{req\_slots}$ using the \textsc{EstimateSlots} function.  
If the locally available timeslot resources are insufficient, the Leader requests the missing slots from the Root node.  
This step leverages hierarchical resource management to elastically expand Leader capacity without distributed negotiation.  
If the Root denies the request or fails to respond within a predefined timeout, the algorithm terminates early to avoid partial or infeasible allocations.  

In Stage~2, the Leader evaluates whether its current domain contains a sufficient number of nodes capable of satisfying the task’s required capabilities $T.C_{req}$.  
If the number of capable nodes is below the minimum required by the task, the Leader initiates on-demand recruitment of mobile resources.
The Leader identifies missing capabilities and broadcasts a task-specific beacon to attract mobile nodes that can provide the required services.  
Candidate mobile nodes responding within the recruitment window are collected and evaluated.  
Selected mobile nodes are granted access and dynamically integrated into the Leader’s domain, expanding the available service pool.  

In Stage~3, the Leader selects an optimal subset of capable nodes based on task semantics, node state, and scheduling impact.  
Timeslots are then allocated from the local resource pool and assigned to the selected nodes.  
Task execution instructions and scheduling information are disseminated to the participating nodes.  
If sufficient communication resources or service capabilities cannot be secured after recruitment, the algorithm terminates with failure.  

Algorithm~\ref{alg:leader_decision} operationalizes the core design of \monaas by decoupling communication resource provisioning from service capability recruitment.  
It prioritizes task semantics over connectivity preservation and enables elastic integration of mobile resources under Leader-level control.  

\subsection{Message Flow of Mobile Resource Aware Protocol}
\label{sec:flow}

\begin{figure}[t]
    \centerline{\includegraphics[width=\columnwidth]{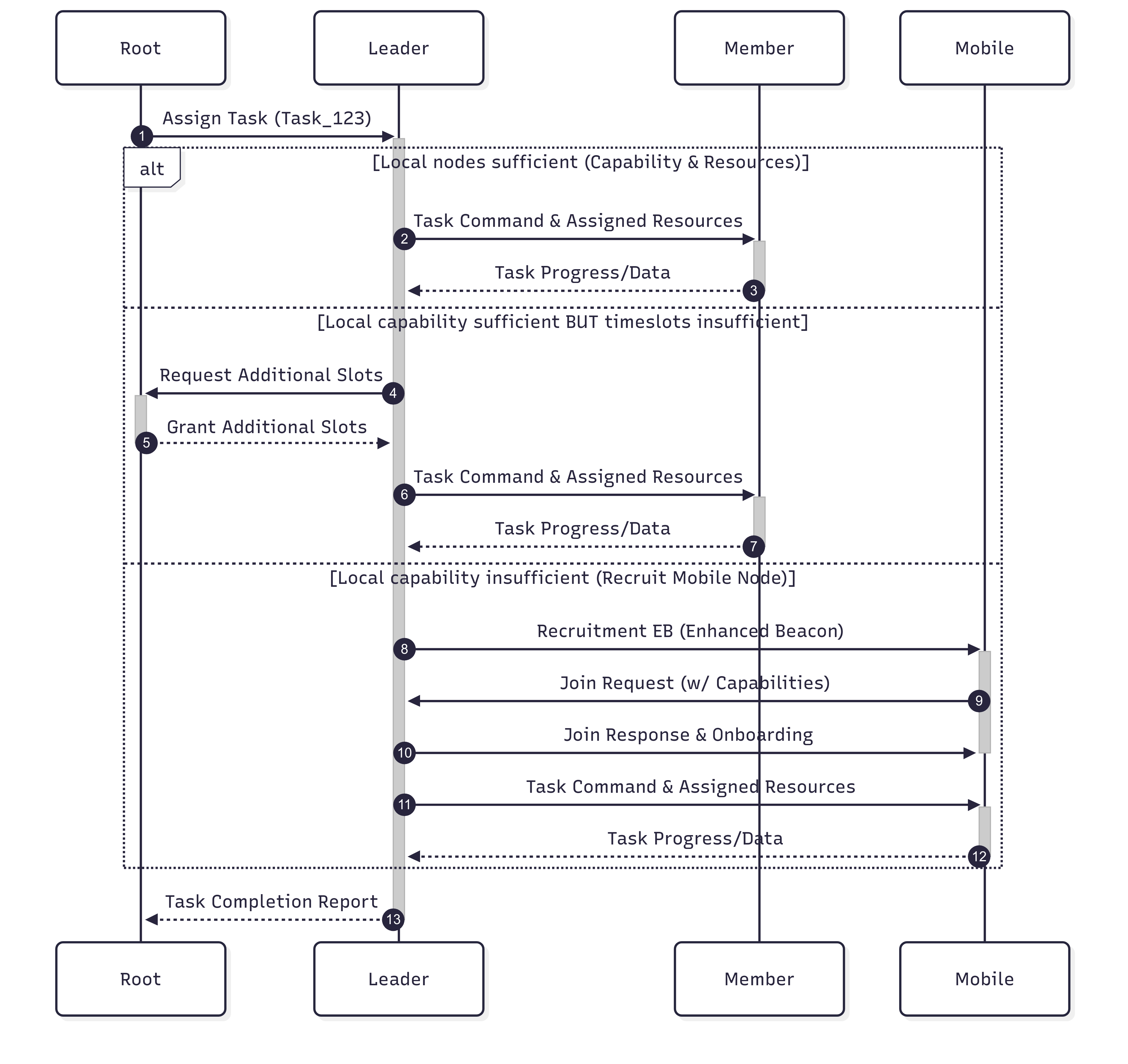}}
    \caption{
        Message workflow of the \monaas protocol showing task dissemination, hierarchical timeslot provisioning, and on-demand mobile resource recruitment.  
        The workflow corresponds to the Leader-side decision logic in Alg.~\ref{alg:leader_decision}.
    }
    \label{fig:workflow}
\end{figure}

Figure~\ref{fig:workflow} presents the end-to-end message exchanging workflow of the \monaas protocol.
The workflow corresponds directly to the Leader-side decision logic presented in Algorithm~\ref{alg:leader_decision} and the timeslot demand estimation given by the $\mathit{req\_slots}$ model.

The workflow begins when the Root assigns a task $T$ to a designated Leader.
Upon receiving the task, the Leader computes the required number of timeslots $\mathit{req\_slots}$ using the \textsc{EstimateSlots} function defined in the task-driven scheduling model.
This computation determines whether the task demand can be satisfied using the Leader’s current local resource pool $R_l(t)$.

If both communication resources and service capabilities are sufficient, corresponding to the success path in Stage~3 of Algorithm~\ref{alg:leader_decision}, the Leader directly issues task commands and assigned TSCH resources to selected Member nodes.  
During execution, Member nodes periodically report task progress and sensed data back to the Leader.  

If service capabilities are sufficient but local timeslot resources are insufficient, the Leader follows the resource expansion path in Stage~1 of Algorithm~\ref{alg:leader_decision}.  
In this case, the Leader requests additional timeslots from the Root based on the deficit between $\mathit{req\_slots}$ and $|R_l(t)|$.  
Upon receiving granted resources, the Leader updates its local resource pool and proceeds with task allocation.  

If local service capabilities are insufficient, the Leader enters the recruitment path described in Stage~2 of Algorithm~\ref{alg:leader_decision}.  
The Leader broadcasts a recruitment message using enhanced beacons that encode task requirements and missing capabilities derived from $T.C_{req}$.  
Mobile nodes evaluate their capability sets against the task requirements and respond with join requests that include capability descriptions.  
Selected mobile nodes are onboarded and dynamically integrated into the Leader’s domain $D_l(t)$.  

After integration, mobile nodes are treated as regular task executors and are assigned TSCH resources from the updated resource pool $R_l(t)$.  
Task execution proceeds with both static and mobile nodes reporting progress and data to the Leader.  

Upon task completion, the Leader aggregates execution results and submits a task completion report to the Root.  
Through this tightly coupled interaction between scheduling logic, resource management, and protocol messaging, \monaas ensures that high-level task semantics are consistently enforced at the MAC layer. 

\begin{table*}[t]
\centering
\caption{Frame Types and Corresponding Subtypes}
\label{tab:frame_types}
\begin{tabular}{cll}
\hline
\textbf{Frame Type} & \textbf{Value} & \textbf{Subtype Semantics} \\
\hline
Beacon & 0x00 &  
[16 bits] Task ID, 
[8 bits] Required Capabilities (bit vector mapping), \\
& & 
[2 bits] Priority Level,
[8 bits] QoS Constraints (percentage),  \\
& & 
[8 bits] Execution Zone, 
[16 bits] Time Window (in timeslots), \\
& & 
[64 bits] Access Credentials, 
[16 bits] Resource Estimate (in timeslots) \\
\hline
Data & 0x01 & 
0x01: Sensor data; 
0x02: Node status; 
0x03: Event data; 
0x04: Location data \\
\hline
Command & 0x02 &
0x01: Schedule update; 
0x02: Join request; 
0x05: Acknowledgment; 
0x06: Activation; \\
 & & 
0x08: MAC test; 
0x09: Test response; \\
 & & 
0x10: Task request; 
0x11: Task response; 
0x12: Task progress; 
0x13: Task completion; \\
 & & 
0x14: Resource request; 
0x15: Resource response \\
\hline
Acknowledgment & 0x03 & No subtype \\
\hline
Event & 0x04 &
0x01: State change; 
0x02: Error; 
0x03: Topology change \\
\hline
\end{tabular}
\end{table*}

The message format of the \monaas is shown in Tab.~\ref{tab:frame_types}.
In the message workflow, the \textit{Assign Task} message sent from the Root to a Leader is encoded as a Command frame (0x02) with subtype 0x10 (Task request), carrying the task identifier and high-level task parameters.  
Subsequent \textit{Task Command \& Assigned Resources} messages from the Leader to Member or Mobile nodes also use Command frames with subtype 0x10, embedding execution parameters and allocated timeslot information.  
During execution, \textit{Task Progress / Data} messages are reported using Data frames (0x01) with subtypes corresponding to sensor data (0x01), node status (0x02), or event data (0x03).  
When local communication resources are insufficient, the Leader issues a \textit{Request Additional Slots} message to the Root using a Command frame with subtype 0x14 (Resource request), and the Root responds with a Command frame of subtype 0x15 (Resource response) to grant additional timeslots.  
If local service capabilities are insufficient, the Leader broadcasts a \textit{Recruitment EB} encoded as a Beacon frame (0x00) carrying a Recruitment Information Element that specifies task ID, required capabilities, priority, QoS constraints, execution zone, time window, access credentials, and resource estimates.  
Mobile nodes respond with a \textit{Join Request} message encoded as a Command frame with subtype 0x02, including their capability descriptors, and successful onboarding is confirmed through a Command frame with subtype 0x05 (Acknowledgment).  
Finally, upon task completion, the Leader reports execution results to the Root using a Command frame with subtype 0x13 (Task completion), completing the protocol message exchange defined in Table~I.

\subsection{Practical Example Using \monaas}
\label{sec:example}

To illustrate the end-to-end operation of \monaas, we walk through a practical example in the factory environment shown in Fig.~\ref{fig:deployment_scenario}.  

\subsubsection{Phase 1: Critical Task Generation}

Consider a sudden chemical leakage detected in Zone~A, which triggers an emergency alert at the factory control center acting as the Root node.  
The Root formulates a high-priority task $T_{\mathit{leak}}$ to assess the situation and contain potential risks.  

The task is defined as follows.  
\begin{itemize}
    \item $id = \texttt{leak\_scan\_A\_01}$.
    \item $P = \texttt{Critical}$.
    \item $Q = \{ \mathit{lat}_{\max} = 200\,\mathrm{ms}, \mathit{pdr}_{\min} = 0.9 \}$.
    \item $C_{req} = \{\mathrm{gas\_sensor}, \mathrm{hd\_camera}\}$.
    \item $Z_{target} = \text{Zone A}$.
    \item $W_{target} = [t_{\mathit{now}}, t_{\mathit{now}} + 300\,\mathrm{s}]$.
\end{itemize}

The Root assigns this task to the Leader responsible for Zone~A, denoted as Leader~A, using a \lstinline{CMD_AssignTask} message.  

\subsubsection{Phase 2: Leader-Side Scheduling Decision}

Upon receiving $T_{\mathit{leak}}$, Leader~A executes the decision logic defined in Alg.~\ref{alg:leader_decision}.  

Leader~A first computes the required number of timeslots $\mathit{req\_slots}$ using the \lstinline{EstimateSlots} function.  
In this scenario, the task generates a data stream with a packet rate $r_{T_{id}} = 2$~packets/s.  
The TSCH slotframe duration is $T_{sf} = 2.02\,\mathrm{s}$.  
The estimated link delivery probability is $L_{\mathit{est}} = 0.8$.
The task priority is \texttt{Critical}, yielding a priority scaling factor $f_{P} = 4$.

Using the model defined in Section~III-B, the required number of timeslots per slotframe is calculated as
\[
\mathit{req\_slots}
= \left\lceil 2 \cdot 2.02 \cdot \frac{0.9}{0.8} \cdot 4 \right\rceil
= \lceil 18.18 \rceil
= 19 .
\]
Leader~A compares this demand against its local resource pool which has 8 available timeslots.
Following Stage~1 of Alg.~\ref{alg:leader_decision}, Leader~A requests 11 additional timeslots from the Root.
The Root allocates the requested resources from the global pool, expanding $R_l(t)$ accordingly.  

Leader~A then evaluates service capability availability within its local domain.  
All static Member nodes in Zone~A are found to provide only basic environmental sensing and lack the required gas sensing and high-definition imaging capabilities.  
This triggers the mobile resource recruitment process described in Stage~2 of Algorithm~\ref{alg:leader_decision}.  

\subsubsection{Phase 3: Mobile Node Recruitment and Integration}

Leader~A embeds the missing capability requirements $C_{req}$ into an Information Element carried by its Enhanced Beacons.  
A nearby mobile Automated Guided Vehicle, denoted as AGV--07, receives the beacon while listening for synchronization.  
AGV--07 advertises a capability set $C_{\mathit{AGV07}} = \{\mathrm{gas\_sensor}, \mathrm{hd\_camera}, \mathrm{manipulator\_arm}\}$.  
The AGV’s operational state indicates sufficient residual energy with a battery level of $85\%$.  

AGV--07 sends a \lstinline{REQ_Join} message to Leader~A to request association.  
Leader~A evaluates the request, confirms capability compatibility and node health, and admits AGV--07 into its domain using a \lstinline{RSP_Join} message.  
As a result, AGV--07 becomes a temporary member of $D_l(t)$ and is included in subsequent scheduling decisions.  

\subsubsection{Phase 4: Task Execution and Reporting}

With both communication resources and service capabilities secured, Leader~A assigns a concrete sub-task to AGV--07 using a \lstinline{CMD_ExecuteSubTask} message.  
Dedicated TSCH timeslots are allocated to AGV--07 from the updated resource pool $R_l(t)$.  
AGV--07 navigates to the target location in Zone~A, performs gas sensing and visual inspection, and reports measurement data during its assigned transmission opportunities.  

Upon completion, Leader~A aggregates the reported data and determines that the task has been successfully executed.  
A final \lstinline{REPORT_TaskStatus} message is sent to the Root, completing the task lifecycle.  
This example demonstrates how \monaas integrates task semantics, hierarchical resource management, and on-demand mobile services into a unified and deterministic scheduling workflow. 

In a conventional 6TiSCH network, the same scenario would proceed in a fundamentally different manner.
AGV-07 would first join the network through standard neighbor discovery and 6P-based negotiation, then rely on application-layer mechanisms to expose its sensing capabilities, and finally be scheduled using traffic-driven resource allocation without explicit awareness of task priority or service semantics.
This multi-stage process introduces significant latency and offers no assurance that the AGV’s specialized capabilities, such as gas sensing and high-definition imaging, would be identified or effectively utilized during an emergency.
In contrast, \monaas enables rapid integration through explicit capability matching, zone-aware recruitment, and task-priority-driven scheduling, transforming mobile nodes from passive connectivity endpoints into active, on-demand service providers.

\subsection{Symbol Notion of \monaas}

Tab.~\ref{tab:notation} shows all symbol notions defined in \monaas for reference.

\begin{table}[t]
\centering
\caption{Notation Summary for the \monaas Framework}
\label{tab:notation}
\begin{tabular}{ll}
\hline
\textbf{Symbol} & \textbf{Description} \\
\hline
$N$ & Set of all nodes in the \monaas network \\
$R$ & Root node responsible for global coordination \\
$L$ & Set of Leader nodes \\
$M$ & Set of static Member nodes \\
$Mo$ & Set of Mobile nodes \\
$n$ & A generic node, $n \in N$ \\
$l$ & A Leader node, $l \in L$ \\
$D_l(t)$ & Local domain managed by Leader $l$ at time $t$ \\
\hline
$T$ & A task instance in \monaas \\
$id$ & Unique identifier of a task \\
$P$ & Task priority level \\
$Q$ & Task QoS requirements \\
$lat_{max}$ & Maximum allowable task latency \\
$pdr_{min}$ & Minimum required packet delivery ratio \\
$C_{req}$ & Capability set required by a task \\
$Z_{target}$ & Target execution zone of a task \\
$W_{target}$ & Execution time window of a task \\
\hline
$C_n$ & Capability set provided by node $n$ \\
$S_n$ & Real-time operational state of node $n$ \\
$b_n$ & Remaining battery energy of node $n$ \\
$load_n$ & Packet queue length of node $n$ \\
\hline
$r_{T_{id}}$ & Task-specific packet generation rate (packets/s) \\
$T_{sf}$ & TSCH slotframe duration \\
$\mathit{pkt\_sf}$ & Packets generated per slotframe ($r_{T_{id}} \cdot T_{sf}$) \\
$L_{\mathit{est}}$ & Estimated link delivery probability \\
$f_P$ & Priority scaling factor derived from task priority $P$ \\
$\mathit{req\_slots}$ & Required number of timeslots for executing task $T$ \\
\hline
$R_{\mathit{global}}$ & Global timeslot resource pool managed by the Root \\
$R_l(t)$ & Local timeslot resource pool of Leader $l$ at time $t$ \\
$R_{\mathit{base}}(l)$ & Baseline timeslots pre-allocated to Leader $l$ \\
$R_{\mathit{alloc}}(l,t)$ & Additional timeslots dynamically granted at time $t$ \\
\hline
\end{tabular}
\end{table}

\section{Experimental Setup and Configuration}
\label{sec:setup}

To rigorously evaluate the MAC-layer performance of the \monaas architecture on real hardware, we use Nordic nRF52840 DK development kits as wireless nodes to conduct experiments.  
Each node ran firmware based on the OpenWSN~\footnote{http://www.openwsn.org/} stack and utilized its mature TSCH MAC-layer implementation.  
The control logic for \monaas and all baseline schemes was implemented directly on top of TSCH, and standard network-layer protocols were disabled to isolate the impact of MAC-layer scheduling.
The slotframe length of TSCH configuration is set to 11 with 20~ms slotduration, resulting a 2.02s slotframe cycle.
By default, each node has a static 1 packet per second rate base.
Actual sending rate will vary under different experiment scenarios, which is explained in following section.

The testbed was deployed in a typical indoor laboratory environment covering approximately 80~m$^2$, with realistic radio-frequency interference from surrounding equipment.
The default transmission power of these nodes is set to 0~dbm.
The experimental network consisted of 11 nRF52840 DK nodes, forming a concise configuration sufficient to validate the core mechanisms of the proposed architecture.

During experiments, the Root node forwarded all received packets to a host PC via a serial interface.  
A Python-based data collection and analysis tool running on the PC parsed the logs in real time and computed all evaluation metrics.  

The network topology and mobility patterns were designed to highlight the contrast between service-centric and connectivity-centric integration paradigms.  
The test network comprised 11 nodes organized into a two-domain topology with a centrally placed Root and two Leader nodes forming partially overlapping communication domains.  
Five static Member nodes were deployed within these domains, establishing a two-level association structure between Members, Leaders, and the Root.  
Three Mobile nodes were initially idle and reserved for on-demand service recruitment.  

Each Mobile node was configured with a distinct service capability profile to create explicit service gaps.  
Mobile-1 provided environmental sensing capabilities for emergency response, Mobile-2 supported visual inspection and surveillance, and Mobile-3 enabled maintenance and actuation tasks.  
In contrast, static Member nodes were limited to basic sensing and communication functions, making advanced task execution dependent on mobile recruitment.
These capabilities were \emph{logically defined abstractions} used to emulate heterogeneous services for protocol evaluation, rather than actual physical sensors or actuators attached to the devices.
 
In the mobility scenario, Mobile-1 was moved from outside network coverage into a Leader’s communication range when a task requiring gas and temperature sensing was issued, triggering service-driven integration.   

\section{Experimental Analysis}
\label{sec:analysis}

This section evaluates the performance of \monaas by comparing it against four representative TSCH baselines spanning static scheduling, distributed adaptation, and centralized control.  
These baselines capture different design assumptions and highlight the fundamental differences between connectivity-centric and service-centric scheduling.  

\begin{itemize}
    \item \textit{Static TSCH:} An idealized reference with an offline, collision-free schedule for a fixed topology and traffic pattern. It provides maximum determinism but cannot adapt to topology, traffic, or mobility changes.
    
    \item \textit{6TiSCH Minimal Configuration~\cite{rfc8180}:} A standard dynamic baseline that uses the 6TOP Protocol to negotiate cells reactively for connectivity restoration. It remains connectivity-driven and unaware of task semantics or mobile service capabilities.
    
    \item \textit{OST~\cite{jeong2020ost}:} A traffic-adaptive scheduler that allocates resources based on local congestion and queue state. It improves flexibility over 6TiSCH Minimal but remains task-agnostic.
    
    \item \textit{FTS-SDN~\cite{bello2020sdn}:} A centralized, flow-based approach where a PC-based SDN controller computes and installs global schedules. Unlike this monolithic control loop, \monaas adopts a hierarchical Controller--Leader architecture that preserves local agility and exploits mobility as a service resource.
\end{itemize}

These baselines provide a comprehensive yardstick for evaluating determinism, adaptability, and control overhead.  
The following analysis relates the observed performance trends directly to these architectural differences.  

\subsection{Scenario 1: The Minimal Cost of Dynamic Service Capabilities}
This baseline scenario addresses the fundamental question: *Can we afford service-oriented capabilities?* Under stable conditions, Static TSCH naturally achieved optimal performance (98.5\% PDR, 270ms latency, 0.8\% RDC)—the theoretical ceiling for rigid approaches. The critical finding is that \monaas achieved virtually identical performance (98.2\% PDR, 350ms latency) while acquiring complete service orchestration capabilities for a mere 0.3\% energy premium (1.1\% vs 0.8\% RDC). This near-negligible cost proves that hierarchical service coordination can be obtained without sacrificing baseline efficiency, establishing dynamic capabilities as an obvious architectural investment.

\begin{figure}[htbp]
\centering
\includegraphics[width=\columnwidth]{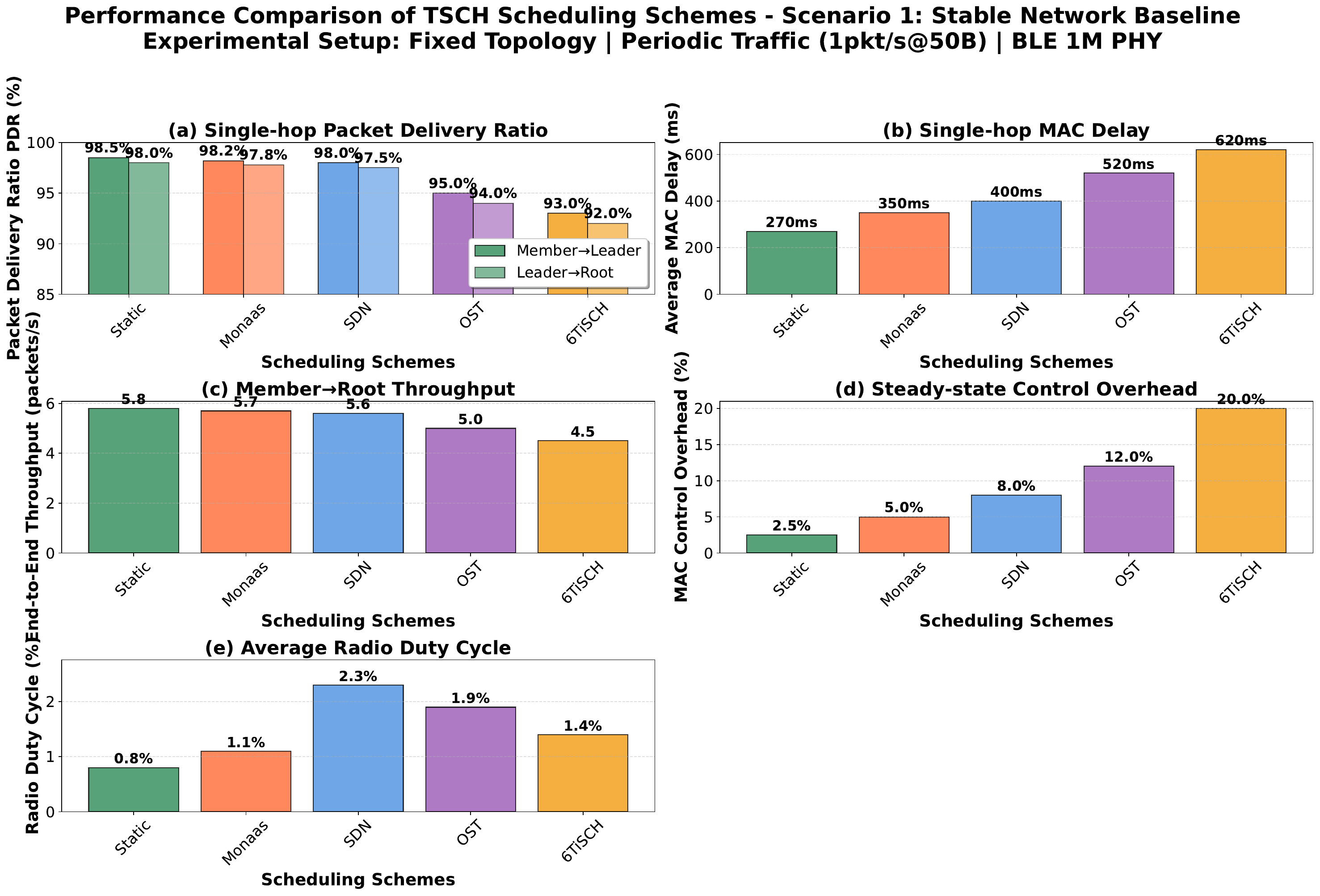}
\caption{Scenario 1: Baseline performance and overhead comparison. (a) Hop PDR. (b) Average hop MAC latency. (c) End-to-end throughput. (d) MAC control overhead. (e) RDC.
}
\label{fig:scenario1}
\end{figure}

The energy cost analysis reveals dramatic differences in coordination strategies across the five approaches. While \monaas pays only 1.1\% RDC through efficient localized Leader beacons and member state synchronization, alternative dynamic schemes prove prohibitively expensive: SDN consumes 2.3\% (2.9× higher) maintaining persistent controller connectivity and global flow consistency, OST expends 1.9\% (2.4× higher) on continuous traffic-driven negotiations, and 6TiSCH requires 1.4\% (1.8× higher) for distributed cell management. The 5.7 packets/s throughput achieved by \monaas with its modest overhead closely matches Static's 5.8 packets/s, while SDN achieves 5.6 packets/s at dramatically higher energy cost. This efficiency advantage creates the crucial energy budget for \monaas's advanced capabilities—task semantic processing, mobile resource integration, capability matching—that prove essential under dynamic stress. Control-plane signaling overhead exhibits the same ordering (Static < \monaas < SDN < OST < 6TiSCH), reflecting that localized hierarchical coordination avoids persistent controller heartbeats and peer-to-peer negotiation storms (see Fig.~\ref{fig:scenario1}).

\subsection{Scenario 2: The High Value of Task-Semantic Capabilities}
Building on Scenario 1's finding that service-oriented capabilities come at a minimal cost, this scenario addresses the complementary question: *Why choose service-oriented approaches?* Under harsh conditions—bursty high-priority tasks combined with link degradation—the results on our testbed provide strong evidence of \monaas's task semantic awareness. We observed that \monaas achieved 99.5\% Task Completion Rate (TCR) while connectivity-centric approaches fell to $\le 35\%$, suggesting that task-driven paradigms can be crucial under stress.

The performance hierarchy further exposes architectural differences in handling dynamics: \monaas (99.5\% TCR, 85\,ms high-priority delay, 92.5\% PDR) performs best by translating task requirements into immediate allocation of contention-free paths (consistent with the EstimateSlots(T) mechanism in Section~III). SDN reaches 96.0\% TCR with 120\,ms delay but treats priorities as generic flows; OST and 6TiSCH, though traffic-aware, are task-semantic-blind (88.0\%/180\,ms and 85.0\%/250\,ms, respectively); Static degrades severely (35.0\% TCR, 2200\,ms), likely because its fixed capacity cannot adapt to a $5\times$ traffic burst.

Across classic metrics, we also observed that \monaas sustains resilience under stress: it maintains 92.5\% PDR (vs. Static 65.2\%, others 75--82\%) and near-ideal deadline performance (DMR $\approx$ 98\% vs. $\le$ 80\% for traffic-driven baselines); average MAC delay stays at 480\,ms (vs. Static 1850\,ms); end-to-end throughput reaches 5.1\,pkts/s (vs. Static 2.1\,pkts/s); energy remains at 1.8\% RDC (vs. SDN 4.2\%, 6TiSCH 2.7\%); and coordination overhead is controlled at 11.5\% (vs. 6TiSCH 28.0\%). Overall, these results indicate that the modest investment established in Scenario 1 can yield substantial benefits when the network is under stress.

Taken together, the $28\times$ delay gap and large TCR differences illustrate the potential benefits of task-semantic awareness under stress on our testbed, while pointing to promising directions for broader validation.

\begin{figure}[htbp]
\centering
\includegraphics[width=\columnwidth]{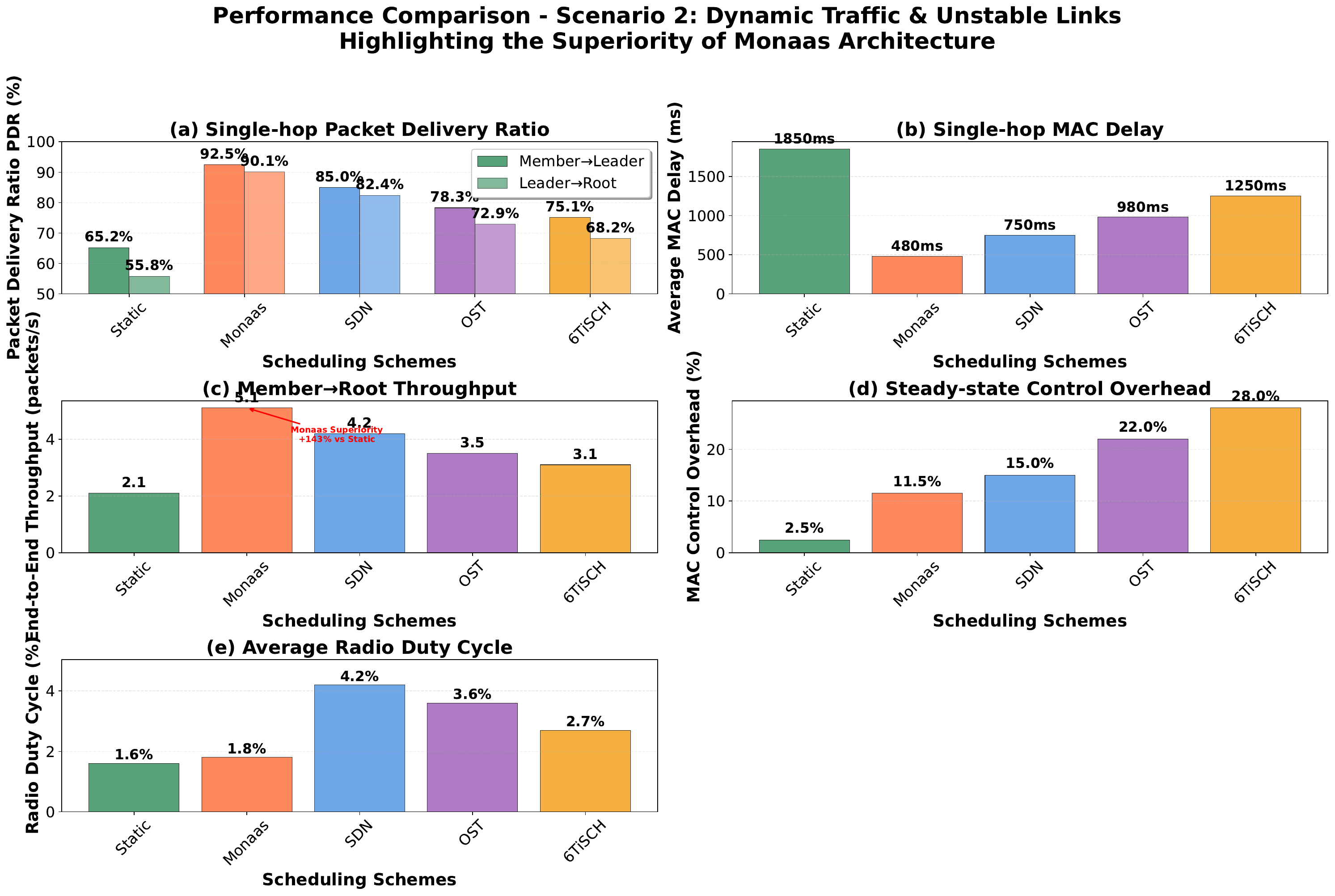}
\caption{
    Scenario 2: Overall network performance under dynamic conditions.
    (a) Hop PDR.
    (b) Hop MAC Latency.
    (c) End-to-end Throughput.
    (d) Control Overhead.
    (e) Radio Duty Cycle.
}
\label{fig:scenario2a}
\end{figure}

\begin{figure}[htbp]
  \centering
\includegraphics[width=\columnwidth]{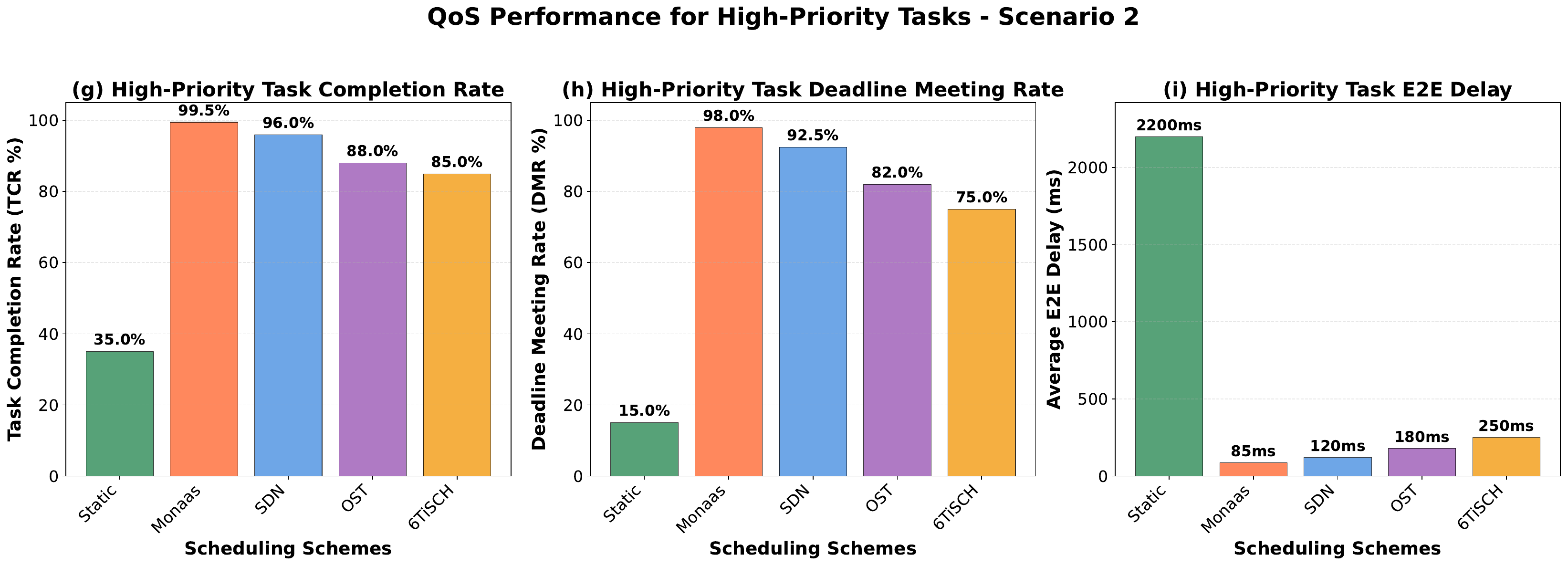}
\caption{
    Scenario 2: High-priority task QoS analysis. 
    (g) TCR. 
    (h) DMR. 
    (i) End-to-end latency for high-priority task.
}
\label{fig:scenario2b}
\end{figure}

\subsection{Scenario 3: On-demand Mobile Resource Integration}
This scenario tests the third core innovation: service-centric mobile resource integration versus connectivity-centric network joining. On our testbed, \monaas achieved a service activation delay of 1.2 seconds compared to SDN (3.5 seconds) and OST/6TiSCH ($\ge$5.8 seconds). More significantly, \monaas demonstrated task-oriented integration (95\% success rate) that can direct mobile nodes to perform specific tasks, while OST/6TiSCH achieved only connectivity-oriented joining (about 30\% success rate) without task-specific command capability.

\begin{figure}[htbp]
\centering
\includegraphics[width=\columnwidth]{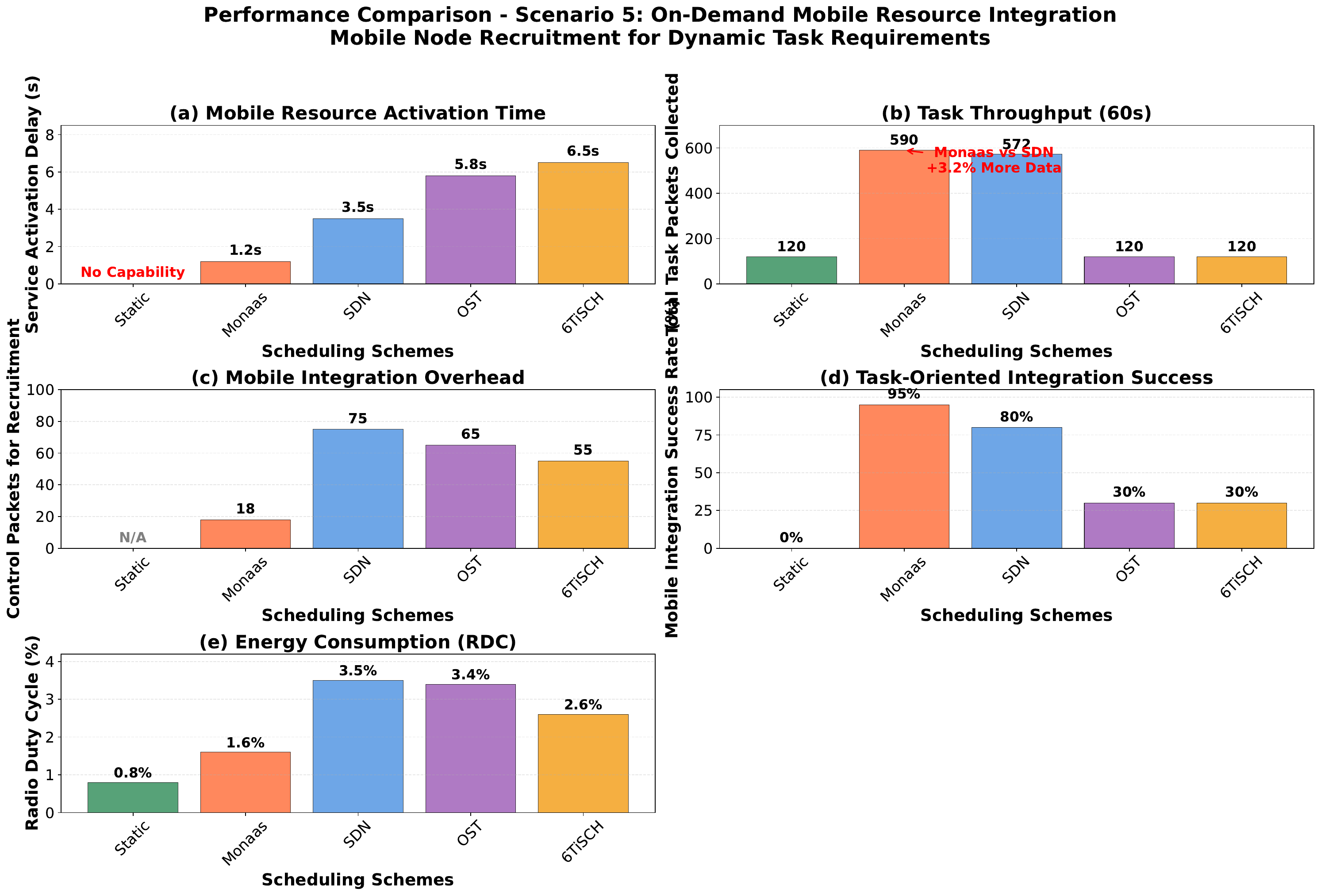}
\caption{
    Scenario 3: Performance of mobile resource integration.
}
\label{fig:scenario4}
\end{figure}

The observed performance differences appear to reflect the paradigm contrast between service-oriented recruitment and connectivity-centric protocols. \monaas's service-oriented model uses Leaders to broadcast Enhanced Beacons containing explicit task requirements, enabling mobile nodes to assess capability matches before attempting integration. This localized, task-aware approach appears to close the loop from capability gap detection to service resource arrival efficiently. In contrast, connectivity-centric approaches use generic protocols: SDN follows an edge-cloud-edge path that may explain the 3.5-second delay, while OST/6TiSCH provide standard network joining without understanding task semantics—they can establish connectivity but cannot direct specific application behaviors. The results suggest that treating mobile nodes as service resources rather than connectivity endpoints may enable more effective dynamic integration (see Fig.~\ref{fig:scenario4}).

\subsection{Scenario 4: Architectural Resilience Under Critical Task Injection}
This scenario tests the comprehensive integration of \monaas's three core innovations under maximum stress: a critical task injection at t=4s that demands immediate resource reallocation. The results suggest that \monaas's service-oriented coordination model may provide superior resilience compared to connectivity-centric architectures. \monaas achieved task activation in 2.5 seconds with minimal service disruption (existing task PDR dropped by only 8\%), while connectivity-centric approaches appeared to struggle: SDN required 6.0 seconds with significant network-wide disruption (PDR dropped to $\sim$76\%), OST/6TiSCH took 7.0-10.0 seconds with severe existing service impact (PDR dropped to 59-71\%).

\begin{figure}[htbp]
\centering
\includegraphics[width=\columnwidth]{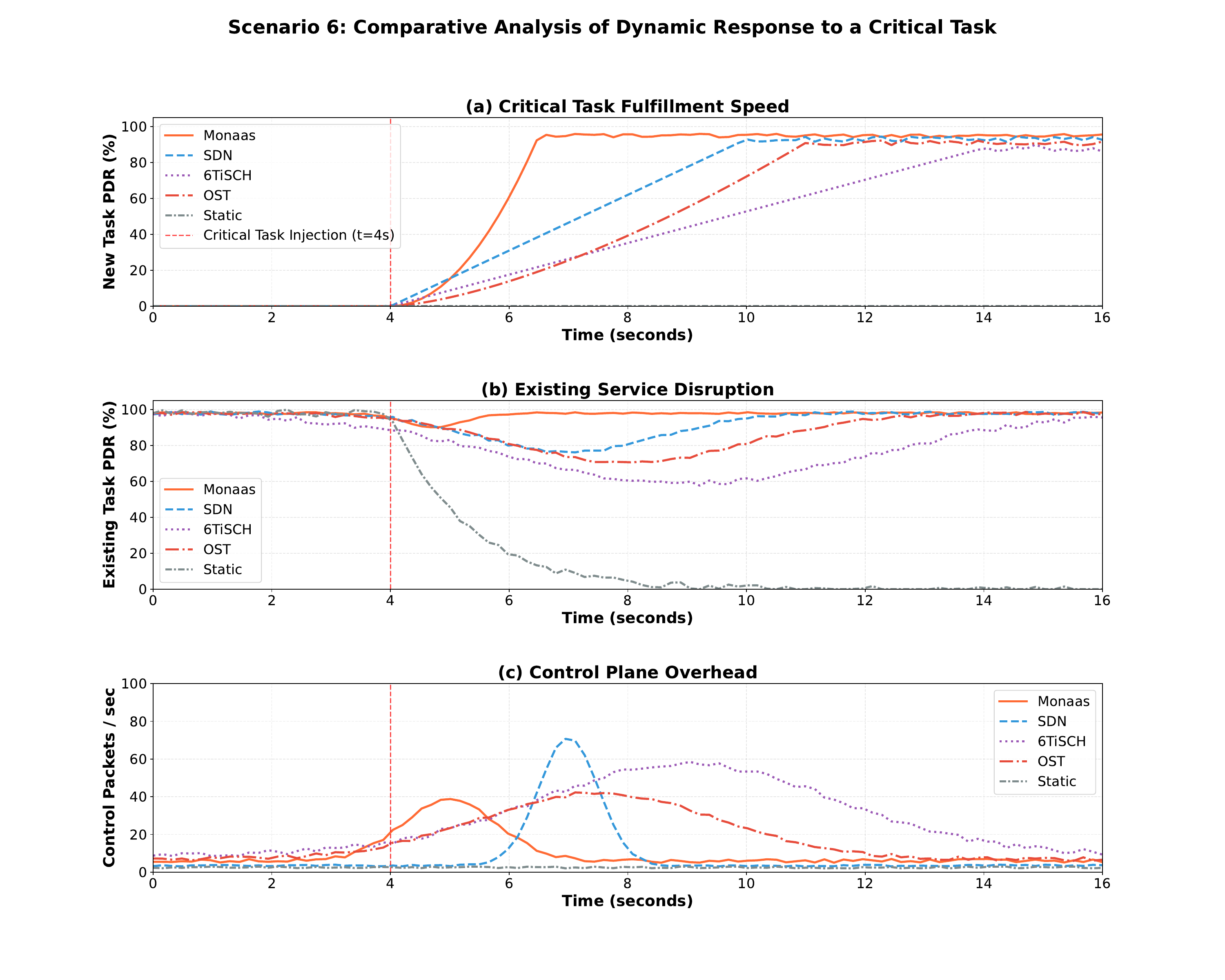}
\caption{
    Scenario 4: Comprehensive comparison of dynamic response and architectural resilience. (a) Response speed to a new critical task. (b) Interference with existing background service. (c) Control plane coordination cost.
}
\label{fig:scenario5}
\end{figure}

The observed performance patterns appear to reflect fundamental architectural differences. \monaas's service-oriented model uses localized Leader-Member coordination that generated a focused control overhead pulse (37 pkts/s peak), suggesting efficient task-semantic resource orchestration. In contrast, connectivity-centric approaches exhibited different coordination costs: SDN's centralized model produced a sharp signaling spike (71 pkts/s), potentially explaining the network-wide disruption, while OST/6TiSCH's distributed negotiation created prolonged overhead that may explain their sustained service interference. The results suggest that treating mobile nodes as elastic service resources rather than network connectivity endpoints could enable more resilient dynamic adaptation (see Fig.~\ref{fig:scenario5}).

\subsection{Experimental Summary}
The experimental results from Scenarios 1 through 5 show that \monaas met or exceeded the performance of all benchmark schemes across multiple metrics, from static efficiency (Scenario 1) to comprehensive robustness (Scenario 5). This indicates that its hierarchical, task-driven architecture possesses certain technical advantages and practical feasibility for addressing the dynamic and heterogeneous application requirements of IIoT.

\section{Limitations and Future Work}
\label{sec:limitation}

While this study validates the fundamental feasibility of the \monaas hierarchical architecture, it has only scratched the surface of the potential encapsulated in the Mobile Node as a Service concept. The current implementation, with its single-layer Leader-Member structure, effectively addresses the basic problem of dynamic resource integration but remains a considerable distance from truly intelligent IoT resource orchestration.

In the current design, node roles are largely fixed after network initialization, which limits the system's adaptability in complex and dynamic environments. In reality, IoT nodes often possess multiple capabilities, serving as executors for some tasks and potentially as coordinators for others. Moreover, the current task-resource matching mechanism is relatively coarse, relying primarily on simple service-type identifiers. True intelligence requires the system to understand the deep semantics of tasks, the multi-dimensional capabilities of resources, and the complex mapping between them.

Our vision for the next generation of \monaas is to transcend these current role limitations by implementing \textbf{dynamic recursive hierarchical management}. The core idea is that any Member node can simultaneously act as a local Leader, recruiting and managing resources below it, with this entire process being transparent to its own superior Leader.

This design would precipitate a qualitative leap forward:
\begin{itemize}[leftmargin=*]
    \item \textbf{Recursive Task Decomposition:} Complex tasks could be automatically decomposed across multiple levels, with each sub-Leader making optimal decisions based on local information.
    \item \textbf{Proximate Resource Integration:} Nodes could proactively recruit the most suitable resources within their communication range, eliminating the need for remote coordination.
    \item \textbf{Decoupling of Management Complexity:} Superior leaders would not need to concern themselves with the internal organization of their subordinates, allowing them to focus on high-level strategy.
\end{itemize}

Although a recursive hierarchical architecture holds great promise, it also introduces new technical challenges: \textbf{multi-level resource conflict arbitration}, \textbf{cross-layer priority propagation}, and \textbf{recursive performance optimization}. Solving these problems will require an interdisciplinary approach, integrating theories from distributed systems, game theory, and artificial intelligence. A deeper challenge lies in achieving true emergent intelligence--enabling the network as a whole to exhibit collective wisdom that surpasses the sum of its individual nodes' capabilities--while maintaining efficient system operation.

A recursive \monaas holds the potential for breakthrough applications in fields such as smart manufacturing, smart cities, and precision agriculture. Imagine a factory floor where each intelligent workstation is not just a production unit but also a micro-workshop manager capable of autonomously recruiting AGVs, sensors, and robotic arms. The entire system would exhibit unprecedented adaptability and collaborative efficiency.

We plan to realize this vision in our future work by first establishing a theoretical framework for the recursive hierarchy, then developing the corresponding protocol mechanisms, and finally validating its effectiveness in real-world application scenarios.

The current \monaas is merely the first cornerstone of this grand blueprint. The true era of Mobile Node as a Service has, perhaps, only just begun.

\bibliographystyle{ieeetran}
\bibliography{liu26monnas}

\end{document}